\documentclass[12pt,preprint]{aastex}

\def\etal{et~al.}
\def\hst{{\it HST}}
\def\vi{\ifmmode(V{-}I)\else$(V{-}I)$\fi}
\def\viz{\ifmmode(V{-}I)_0\else$(V{-}I)_0$\fi}

\def\gz{\ifmmode(g_{475}{-}z_{850})\else$(g_{475}{-}z_{850})$\fi}
\def\gzz{\ifmmode(g_{475}{-}z_{850})_0\else$(g_{475}{-}z_{850})_0$\fi}
\newcommand\lta{\mathrel{\rlap{\lower 3pt\hbox{$\mathchar"218$}}
     \raise 2.0pt\hbox{$\mathchar"13C$}}}
\newcommand\gta{\mathrel{\rlap{\lower 3pt\hbox{$\mathchar"218$}}
     \raise 2.0pt\hbox{$\mathchar"13E$}}}
\def\mbari{\ifmmode\overline{m}_I\else$\overline{m}_I$\fi}
\def\mbarz{\ifmmode\overline{m}_z\else$\overline{m}_z$\fi}
\def\mbar{\ifmmode\overline{m}\else$\overline{m}$\fi}
\def\Mbar{\ifmmode\overline{M}\else$\overline{M}$\fi}
\def\Mbarz{\ifmmode\overline{M_z}\else$\overline{M}_z$\fi}

\slugcomment{2005 ApJ, 625, 121 }

\shorttitle{ACS VCS: SBF Calibration}
\shortauthors{Mei et al.}

\begin{document}


\title{The ACS Virgo Cluster Survey V: SBF Calibration 
for Giant and Dwarf Early-type Galaxies}

\author{Simona Mei\altaffilmark{1}, John P. Blakeslee\altaffilmark{1}, John L. Tonry\altaffilmark{2},  Andr\'es Jord\'an\altaffilmark{3,4},  Eric W. Peng\altaffilmark{5,6}, Patrick C\^ot\'e\altaffilmark{5,6}, Laura Ferrarese\altaffilmark{5,6}, Michael~J.~West\altaffilmark{7}, David~Merritt\altaffilmark{8}, Milo\v s~Milosavljevi\'c~\altaffilmark{9,10}}


\altaffiltext{1}{Department of Physics and Astronomy, Johns Hopkins University, Baltimore, MD 21218; smei@pha.jhu.edu, jpb@pha.jhu.edu}
\altaffiltext{2}{Institute of Astronomy, University of Hawaii, 2680 Woodlawn Drive, Honolulu, HI 96822; jt@ifa.hawaii.edu}
\altaffiltext{3}{European Southern Observatory, Karl--Schwarzschild--Str. 2, 85748 Garching, Germany; ajordan@eso.org}
\altaffiltext{4}{Astrophysics, Denys Wilkinson Building, University of Oxford, 1 Keble Road, Oxford, OX1 3RH, UK}
\altaffiltext{5}{Herzberg Institute of Astrophysics, National Research Council, 5071 West Saanich Road, Victoria, BC, V9E 287, Canada; eric.peng@nrc--cnrc.gc.ca, patrick.cote@nrc--cnrc.gc.ca, laura.ferrarese@nrc--cnrc.gc.ca}
\altaffiltext{6}{Department of Physics and Astronomy, Rutgers University, Piscataway, NJ 08854}
\altaffiltext{7}{Department of Physics and Astronomy, University of Hawaii, Hilo, HI 96720; westm@hawaii.edu}

\altaffiltext{8}{Department of Physics, Rochester Institute of Technology, 54 Lomb Memorial Drive, Rochester, NY 14623; david.merritt@rit.edu}
\altaffiltext{9}{Theoretical Astrophysics, California Institute of Technology, Pasadena, CA 91125; milos@tapir.caltech.edu}
\altaffiltext{10}{Sherman M. Fairchild Fellow}


\begin{abstract}
As part of the Advanced Camera for Survey (ACS) Virgo Cluster Survey, we have 
measured Surface Brightness Fluctuations (SBF) in a sample of 100 early-type
Virgo galaxies. Distances derived from these measurements are needed to 
explore the three-dimensional structure 
of the Virgo Cluster, study the intrinsic parameters of globular clusters
associated with the program galaxies,
and compare with the galaxy distances derived
from globular cluster luminosity functions.
Our SBF measurements have been performed in the F850LP bandpass of the
Wide Field Channel of the ACS on the Hubble Space Telescope.
These are the first measurements of this kind, and 
we present the first SBF calibration for this bandpass.
The measured fluctuations depend on galaxy stellar population variations, which we 
quantify by galaxy color $(g_{475}-z_{850})_0$, where $g_{475}$ and $z_{850}$ are
the galaxy magnitudes, respectively, in the F475W and F850LP ACS filters.
We derive the following calibration
 for the absolute SBF magnitude $\overline M_{850}$: 
\begin{equation} \nonumber
\overline M_{850} =  -2.06 \pm 0.04 + (2.0 \pm 0.2) \times [ (g_{475}-z_{850})_0-1.3 ]  \nonumber
\end{equation}
in the range $1.3 <(g_{475}-z_{850})_0 \le 1.6$, and
\begin{equation} \nonumber
\overline M_{850} =   -2.06 \pm 0.04  +  (0.9 \pm 0.2) \times [ (g_{475}-z_{850})_0-1.3 ] \nonumber
\end{equation}
in the range $1.0 \le(g_{475}-z_{850})_0 \leq 1.3$.
The quoted zero-point uncertainty here includes all sources of internal error;
there is an additional systematic uncertainty of $\sim\,$0.15~mag, due to the 
 uncertainty in the distance scale calibration.

Physically, the two different color regimes correspond to
different galaxy types: giant ellipticals and S0s at
the red end, and early-type dwarfs at the blue end.
For the first time in SBF studies, we are able
to provide a firm empirical calibration of SBF
in early--type dwarf galaxies.
Our results agree with stellar population model predictions from 
Bruzual \& Charlot (2003) in the range $1.3 < (g_{475}-z_{850})_0 \le 1.6$,
while our empirical slope is somewhat steeper than the theoretical prediction
in the range $0.9 \le(g_{475}-z_{850})_0 \leq 1.3$. 

\end{abstract}

\keywords{galaxies: distances and redshifts ---
galaxies: dwarf ---
galaxies: elliptical and lenticular, cD ---
galaxies: clusters: individual (\objectname{Virgo})
}

\def\hst{{\it HST}}

\section{Introduction}

The Advanced Camera for Surveys (ACS; Ford et al. 1998) Virgo Cluster Survey is
a project based on observations with the ACS on the {\it Hubble Space Telescope}
(\hst), aimed at the exploration of the properties of 100 early--type galaxies
in the Virgo Cluster, the study of their globular cluster systems, and the
reconstruction of Virgo's three dimensional structure.  To measure the galaxy
distances needed to realize these goals, we
have used measurements of Surface Brightness Fluctuations (SBF). The SBF method
was first proposed by Tonry \& Schneider (1988) and it is based on the
fact that the variance of normalized poissonian fluctuations of the galaxy
stellar population depends on galaxy distance (for
reviews see Jacoby et al. 1992 and Blakeslee et al. 1999).  At present,
SBF measurements have been used to
determine galaxy distances out to $\approx$~7000~km~s$^{-1}$, using data
from ground--based telescopes and \hst\ (Ajhar \etal\ 1997, 2001; Tonry \etal\ 1997, 2001; Jensen
\etal\ 1999, 2003; Blakeslee \etal\ 2001, 2002; Mei \etal\ 2001, 2003; Liu \etal\
2001, 2002; Mieske et al. 2003; and references in Mei et~al. 2005, hereafter Paper~IV).

For the Virgo Cluster in particular, SBF measurements have been published
by several groups using data from both ground--based telescopes
(Tonry et al.\ 1990, 2000, 2001; Pahre \& Mould 1994; Jensen et al.\ 1998;
Jerjen et al. 2004) and \hst\ (Ajhar et al.\ 1997;
Neilsen \& Tsvetanov 2000; Jensen et al.\ 2003).
West \& Blakeslee (2000) have used published ground-based SBF data in a first
examination of the three-dimensional structure of this cluster, particularly the
cluster's ``principal axis'' and its relations to the surrounding large-scale
structure. 
Our new ACS dataset (for details, see C\^ot\'e et~al. 2004;
Paper~I) constitutes the
largest, most complete, and most homogeneous SBF sample of Virgo galaxies yet
assembled.

Since the absolute magnitude of the SBF varies as a function of the stellar
population age and metallicity, SBF distance measurements in a given
observational bandpass must be carefully calibrated in terms of stellar
population observables, typically the galaxy color.  
Using an extensive ground-based $I$-band SBF survey,
Tonry et al. (1997, 2001) have
established that it is possible to calibrate the SBF method 
using the integrated galaxy colors. 
In particular, they find that for
bright ellipticals: 1) the absolute SBF magnitude $\overline M_I$ is 
a linear function of the galaxy $(V{-}I)_0$ color 
over the range $1 < (V{-}I)_0 < 1.3$ 
with an internal scatter $\lta 0.1$~mag; 
2) the zero-point of this relation is universal over this color range; 
3) the slope of the relation is consistent with theoretical predictions 
from recent stellar population models.  For observational
bands other than the $I$-band, the relation between SBF absolute
magnitude and galaxy colors has to be calibrated in that particular band. 
Ajhar et al. (1997) and Jensen et al. (2003) have calibrated SBF 
absolute magnitudes as a linear function of galaxy color
for use, respectively, with the Wide Field and Planetary Camera-2 (WFPC-2) and
NICMOS on the \hst.

This paper provides the first calibration for SBF using the F850LP
filter with the ACS, i.e, calibration of the absolute F850LP SBF magnitude 
$\overline M_{850}$, as a function of dereddened color in the F475W 
and F850LP filters. 
Observations and SBF reductions have already been
described in Paper~I, Jord\'an et al. (2004; Paper~II) and Paper~IV.
In \S~2 the main steps of the SBF reduction are
briefly summarized. In \S~3 the calibration method is discussed and the
results are presented and analyzed. In \S~4 the theoretical predictions from
Bruzual \& Charlot (2003) stellar population models are discussed. We 
summarize and conclude in \S~5.
The calibration presented here will be used in future papers to derive SBF
distances for the galaxies in our sample and to explore the three-dimensional
structure of the Virgo Cluster.

\section{SBF measurements}

\subsection{Observations and SBF reduction}

Our observations and data reduction procedures have been discussed in 
Paper~I and Paper~II, respectively.
In Paper~IV, we have addressed the specific issues related to SBF measurements
with the ACS. Since the ACS camera presents significant geometrical distortions,
the correction of these distortions involves image resampling by
an interpolation kernel.  For our SBF measurements we use images
geometrically corrected with the {\it drizzle} software (Fruchter \&
Hook 2002) using the Lancsoz3 interpolation kernel.
We have shown in Paper~IV that, when using this kernel, 
the distortion correction does not
add significant error to the SBF distance measurements.
The main steps of SBF measurements are also described in Paper~IV.
We summarize here the main points.

We measure SBF in the F850LP bandpass of the ACS Wide Field Camera.
The process involves fitting a smooth model to
each galaxy in both the F850LP and the F475W band 
(Papers~II and~IV) and subtracting the models
from the original images to produce ``residual images''.
The residual images in each band are then used to construct catalogs
of external sources, i.e. globular clusters and background galaxies.
From these catalogs we create a 'source mask' to remove the contamination
from the sources to the surface brightness variance.
All sources brighter than a completeness magnitude $m_{cut}$,
that corresponds to a completeness of $90\%$ at a given radius, 
are masked from the image.  
The power spectrum of the residuals is then calculated in multiple concentric annuli
centered on the galaxy center, by multiplying successive annular mask functions with the 
'source mask' (i.e., the total mask function). The image power spectrum is the sum of two 
components: a flat, white noise power spectrum and the combined power spectrum 
of the SBF and undetected external sources, both convolved by the PSF (Point Spread Function)
in the spatial domain. In the Fourier domain, this sum can be written as:
\begin{equation}
P(k) = P_0 \times E(k) + P_1 \,,
\label{eq:pk}
\end{equation}
where the ``expectation power spectrum'' $E(k)$ is  the
convolution of the power spectra of the normalized PSF and of the total
mask function of the annular region being analyzed. In Paper~IV 
we have shown
that this power spectrum is not significantly modified by pixel correlation
induced by our interpolation kernel (see Paper~IV for details).
Armed with a clearer understanding of the systematic errors arising from the data
interpolation, we have measured unbiased values of
$P_0$ and $P_1$ for multiple annuli in each galaxy as described in Paper~IV. 
As the minimum innermost
 radius, we have taken $\approx 1.5\arcsec$. 
However when dust is present in the galaxy
center, the dust regions are masked (see Paper~I), and the minimum 
radius without dust contamination has been taken as the
minimum annular radius. 
For the maximum outermost annular radius, we have chosen the radius
where the galaxy brightness falls to half that of the sky (the range
of maximum outermost radii being between $\approx 10\arcsec$ and 
$\approx 40\arcsec$).
The next step is then to extract the SBF signal from $P_0$ by subtracting
off the residual variance $P_r$ from faint sources including undetected
globular clusters and distant background galaxies. Because of the 
exquisite resolution and depth of our ACS images, these correction are
very small (see the following section).
\begin{figure*}
\epsscale{1.}
\centerline{\plotone{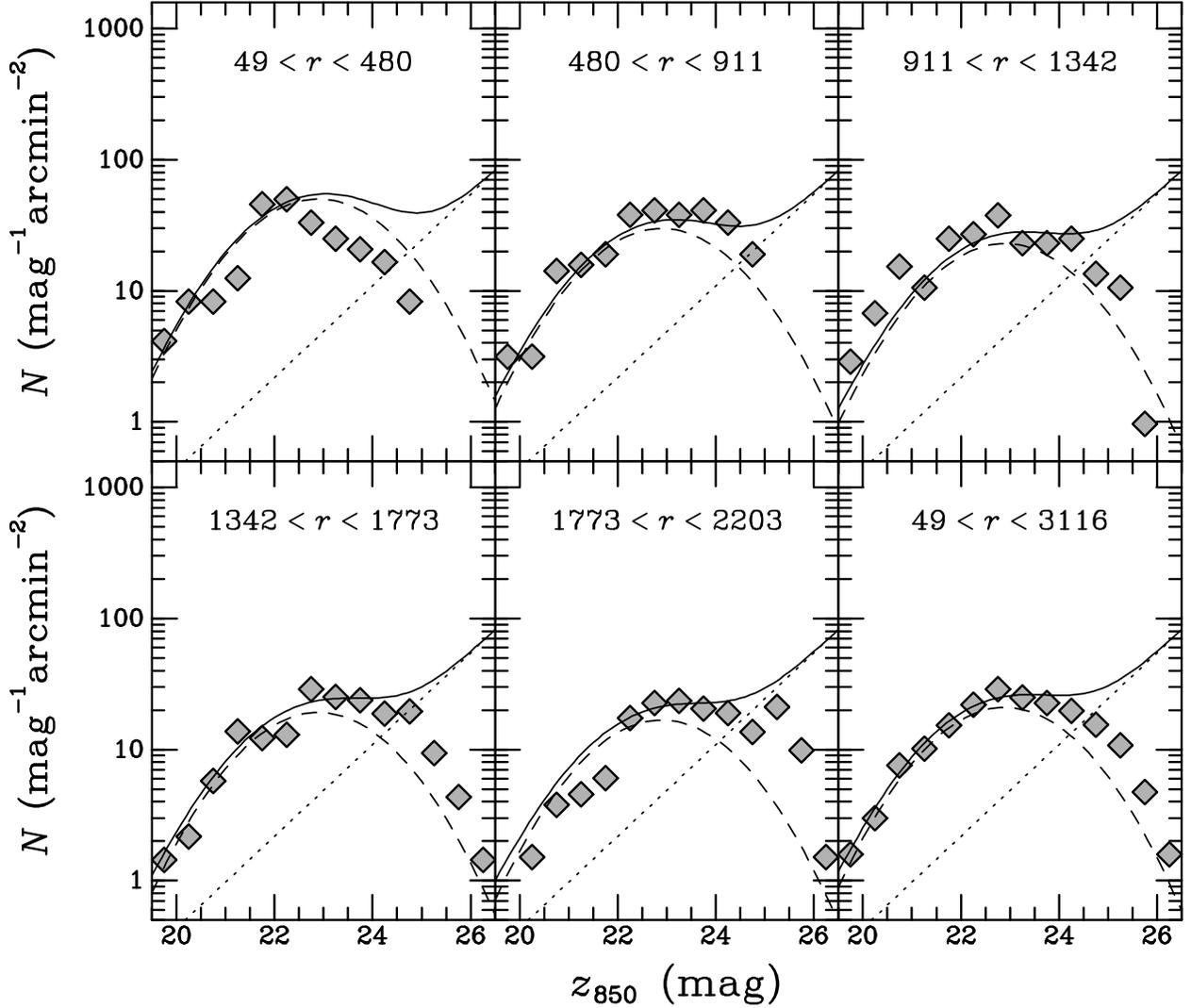}}
\caption {The fit of total external source counts as a function of magnitude
for NGC~4472 (VCC~1226), shown as a continuous line. 
The globular cluster luminosity function fit is shown by the dashed line, and 
the background galaxy luminosity function by the dotted line.
Different plots correspond to different annuli, whose range is shown in the
top of each plot. The diamonds are the total number of external sources in magnitude 
intervals of 0.4~mag.{\label{gc_example}}}
\end{figure*}

\subsection{Contribution to the SBF from external sources}

The contribution to the power spectrum from these external sources is indistinguishable
from the SBF contribution, since both are spatially convolved with
the PSF. Following Tonry \etal\ (1990) and Blakeslee \& Tonry (1995),
we can write 

\begin{equation}
P_{0}=P_{SBF}+P_{r}
\end{equation}
where
\begin{equation}
P_{r}= {\sigma^2_{gc}+\sigma^2_{bg} \over \langle \mbox{\it galaxy} \rangle } \,,
\end{equation}
 $\sigma^2_{gc}$ is the
contribution to the fluctuations due to globular clusters, and
$\sigma^2_{bg}$ is the contribution due to background galaxies. 
The mean galaxy brightness in the denominator is chosen for
compatibility with the $P_0$ definition above.

To estimate the $P_r$ contribution, sources were extracted from the same
catalog used for masking external sources from the residual images 
(see above and Paper~IV).
Source counts were fitted to a composite
globular cluster and external source luminosity function using a
maximum likelihood approach. For the
former, we adopted a Gaussian luminosity function, where $N_{gc}$ is a function
of both radius $r$ (distance from the galaxy center) and magnitude $m$:
\begin{equation}
N_{gc}(m,r) = \frac{N^{0}_{gc}(r)}{\sqrt{2\pi}\sigma}\,
\exp\left({{-(m-m^{gc}_{peak})^2}\over{2\sigma^2}}\right).
\end{equation}
For the Virgo Cluster in the F850LP filter, we expect
$m^{gc}_{peak}  \approx $ 23~mag  and $\sigma \approx 1.35$  
\cite{har91,whi95,fer00a,fer03,pen05}.  
We use these values as our initial estimates
but iterate on $m^{gc}_{peak}$ to improve the fitted
luminosity functions.

For the background galaxies, we assumed a
power-law luminosity function:
\begin{equation}
  N_{bg}(m) = N^{0}_{bg} 10^{\gamma m} 
\end{equation}
with $\gamma = 0.35$.  From the ACS faint galaxy counts of Benitez et al. (2004)  $\gamma = 0.33$ in the F814W filter.
At present, a value for $\gamma$ in F850LP is unavailable in the literature.
However, the F814W and F850LP bandpasses are quite similar so we 
expect no strong differences in $\gamma$ between these filters.
A fit of  faint galaxy counts in our ACS blank fields gives us a slope
of $\gamma = 0.37$ (Peng et al. 2005).
If we vary $\gamma$ in the range 0.33 to 0.37, our fits produce differences
of only a few thousandths of a magnitude in the final SBF magnitudes.

In fitting the total source counts ($N_{ogc}$ and $N_{obg}$), we kept as fixed
parameters $\sigma$ and $\gamma$, and iterated on $m^{gc}_{peak}$,
as a function of the galaxy distance. 
Sources identified 
as bright foreground stars were removed from the images
but not included in the fits.  From the estimated $N_{ogc}$ and
$N_{obg}$ per pixel, $P_{r}$ was calculated as the sum of
\cite{bla95}
\begin{eqnarray}
\sigma^2_{gc}&\,=\,&\frac{1}{2} N_{ogc} 10^{0.8[m_1-m^{gc}_{peak}+0.4\sigma^2\ln(10)]}\\ \nonumber
& \,\times\, & \hbox{erfc}[\frac{m_{cut}-m^{gc}_{peak}+0.8\sigma^2\ln(10)}{\sqrt{2}\sigma}]
\end{eqnarray}
and
\begin{equation}
\sigma^2_{bg}=\frac{N_{obg}}{(0.8-\gamma) 
  \ln(10)}10^{0.8(m_1-m_{cut})+\gamma(m_{cut})}.
\end{equation}
where $m_1$ is the photometric zero-point on the AB system
(24.862 mag for the F850LP; Sirianni et al. 2005) plus reddening
corrections (Schlegel et al. 1998) as given in Paper~II.
An example of the external source luminosity function is shown in
Fig~\ref{gc_example} for NGC~4472 (VCC~1226).
Finally, the $z$-band SBF magnitudes are given by
\begin{equation}
\overline m_{850} =  -2.5 \log(P_0-P_{r}) + m_{1} \label{eq:mag} \,,
\end{equation}

The uncertainty on  $\overline m_{850}$ in each galaxy has been calculated by
propagating the measurement errors in Eq.~\ref{eq:mag} for the different annuli. 
This means that we have added in quadrature the errors due to the uncertainty in
the fit of $P_0$, the uncertainty in the galaxy flux, and the uncertainty in $P_{r}$ 
($\approx$ 25\% $P_{r}$, calculated as the variance in  $P_{r}$ when different
initial fit parameter are chosen).  We neglect the $\sim\,$1\% systematic 
uncertainty in $m_{1}$ (Sirianni et al. 2005) for the calibration,
since it is common to all galaxies.
%
The total uncertainty on the average $\overline m_{850}$ for each galaxy
has been calculated as the quadrature sum of the errors for the different 
annuli which we are averaging, divided by the number of annuli.

Finally, since $\overline m_{850} - \overline M_{850} = 5 log (d)$, where $d$
is the galaxy distance, two more sources of uncertainty must be added 
in quadrature to the measurement error in $\overline m_{850}$
for the purpose of this calibration.
First, we include an expected  ``cosmic scatter'' of 0.05~mag in $\overline M_{850}$
(see Tonry et al.\ 1997), representing the intrinsic dispersion in $\overline M_{850}$
for galaxies at the same observed color.
Second, an uncertainty of 0.11~mag is introduced to account for the actual distance dispersion 
from depth effects in the Virgo Cluster. This value has been derived from the 
rms scatter of $\sim3^\circ$ in position on the sky of the 100 galaxies in our sample.


\begin{figure}
\epsscale{.80}
\centerline{\plotone{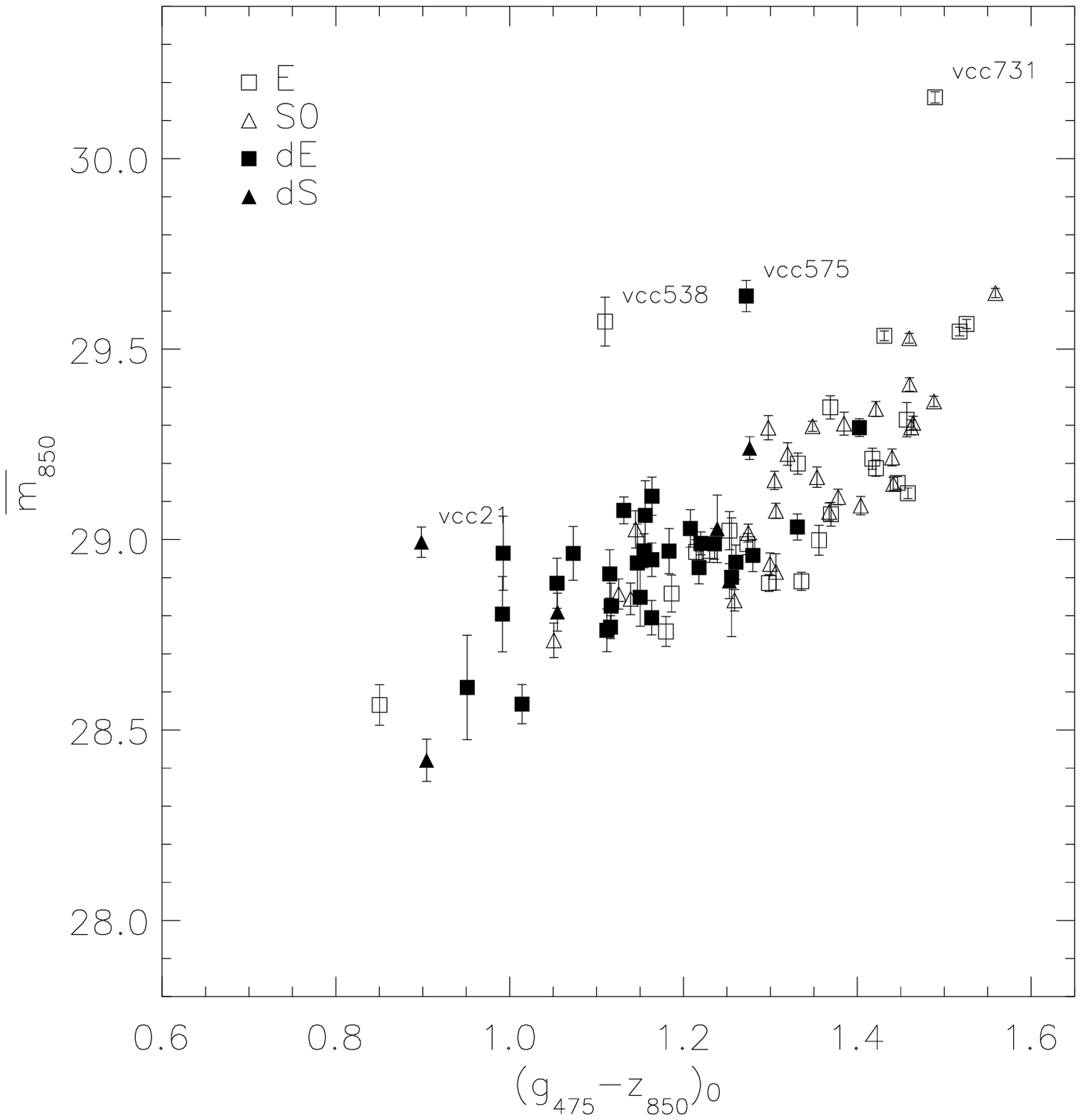}}
\caption{ Our SBF galaxy sample (85 galaxies) is shown.
Elliptical galaxies are represented by open squares, S0 by open triangles, dwarf ellipticals by filled squares, and dwarfs S0 by filled triangles. The outliers are labeled.
{\label{allsample}}}
\end{figure}

\begin{figure}
\epsscale{.80}
\centerline{\plotone{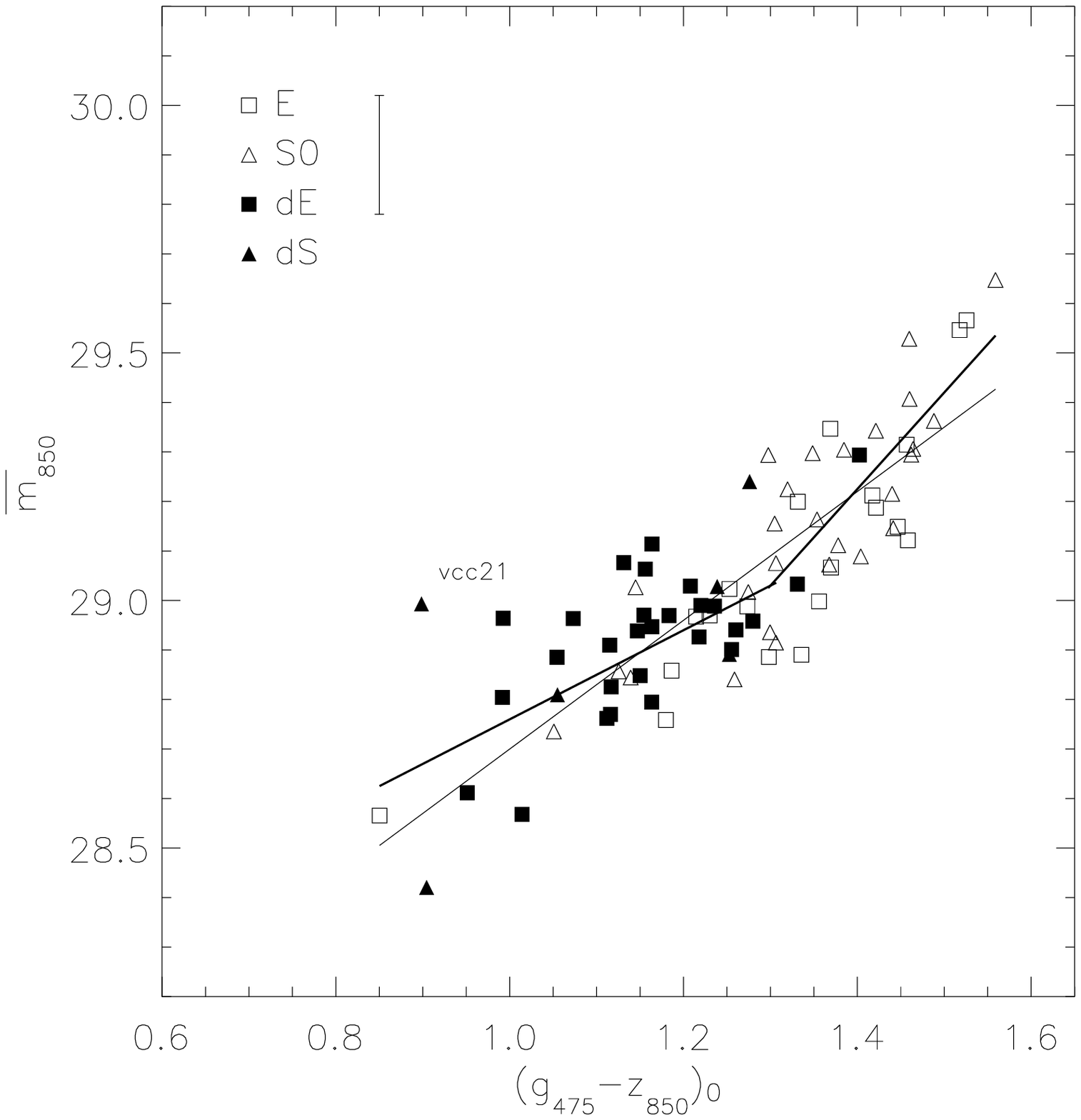}}
\caption {Our final calibration sample (81 galaxies, plus VCC~21; see text) is shown, together with
the fitted relations.
Elliptical galaxies are represented by open squares, S0 by open triangles, dwarf ellipticals by filled squares, and dwarfs S0 by filled triangles.
The straight line is the fit to the full sample, while the thick continuous 
lines are, respectively, fits for the red and blue ends. 
At the top left, the typical
SBF magnitude error (0.12~mag, including the uncertainty due to the Virgo
cluster depth and intrinsic SBF dispersion) is shown. {\label{slope}}}
\end{figure}

\section{Calibration of the ACS $\overline M_{850}$}

\subsection{Color measurements}

We seek to calibrate $\overline M_{850}$ as a function of the 
galaxy color (F475W$-$F850LP), hereafter referred to \gz, 
 since the ACS $F475W$ approximates the SDSS $g$ and F850LP the SDSS $z$. 
The galaxy colors are measured in the same annuli, and with
the same masks, as those used for measuring the SBF.
The sky values in the two filters for our
final calibration sample are the same as in Paper~II.
Galaxies for which the sky estimates were
especially difficult (due to strong gradients from bright neighboring galaxies)
have been excluded for the purposes of this calibration (see below).
The errors on the colors have been calculated by summing in 
quadrature the error in galaxy flux and the error due to sky subtraction.
When the colors have been averaged on more annuli, the errors in each one
have been added in quadrature, squared and divided by number of annuli.
An additional uncertainty of 0.01~mag in color is added in quadrature
as an estimate of the error due to imperfect flatfielding (see, e.g.,
Sirianni \etal\ 2005).

The colors have been dereddened as explained in Paper~II, with the reddening
for each galaxy position from Schlegel et al. (1998). The mean reddening
for the sample is $E(B-V)=0.029$~mag, with a standard deviation of 0.008~mag.
The adopted extinction corrections are appropriate for elliptical galaxies, 
$A_g = 3.634$ $E(B-V)$ and $A_z = 1.485$~$E(B-V)$.
Hereafter, we use \gz $_0$ to refer to the reddening-corrected galaxy color.

\subsection{Calibration Slope}

For the calibration of the $I$-band SBF survey, Tonry et al. (1997) divided
their sample into galaxy groups and clusters, and assumed that, since the galaxies
within an individual group or cluster are at nearly the same distance, SBF magnitudes
will simultaneously fit the relation:
\begin{equation}\label{tonry}
\overline m_I = \langle \overline m^0_I \rangle + \beta [(V-I)_0 - 1.15]
\end{equation}
where $ \langle \overline m^0_I \rangle$ is the group mean value at the
chosen fiducial color $(V-I)_0 = 1.15$~mag, that was fitted for each group
to establish the universality of the SBF calibration.  
To obtain the absolute SBF magnitude $\overline M_I $, Tonry et al.\ (1997)
calibrated the zero-point in Eq.~\ref{tonry} with Cepheid distances
to galaxies in their groups.  Thus,
they obtained the following calibration, valid in the color range 
$1.0 \le (V-I)_0 \le 1.3$:
\begin{equation}
\overline M_I = (-1.74 \pm 0.07) +(4.5 \pm 0.25) [(V-I)_0 - 1.15].
\end{equation}
The Ajhar et al.\ (1997) \hst/WFPC2 sample spans a range in galaxy \vi$_0$ color
between 1.15 and 1.3~mag. Their calibration for the WFPC2 F814W 
bandpass is
\begin{equation}
\overline M_{814} = (-1.73 \pm 0.07) +(6.5 \pm 0.7) [(V-I)_0 - 1.15].
\end{equation}
The Jensen et al. (2003) HST/NICMOS sample in the F160W ($1.6\,\mu$m) filter
was calibrated over the range  $1.05 \le (V-I)_0 \le 1.24$ as:
\begin{equation}
\overline M_{160} = (-4.86 \pm 0.03) +(5.1 \pm 0.5) [(V-I)_0 - 1.16].
\end{equation}

These three samples were heavily weighted
towards bright, red, early--type galaxies. The Ajhar \etal\ and 
Jensen \etal\ samples (due the small field sizes) 
were mainly confined to the central regions of bright galaxies,
which tend to be quite red.  Blakeslee \etal\ (2001) concluded that
this fact explained the steep slope found by Ajhar \etal, since the
stellar population models predict a steepening of the SBF--color
relation for very red populations.

\begin{figure}
\epsscale{.80}
\centerline{\plotone{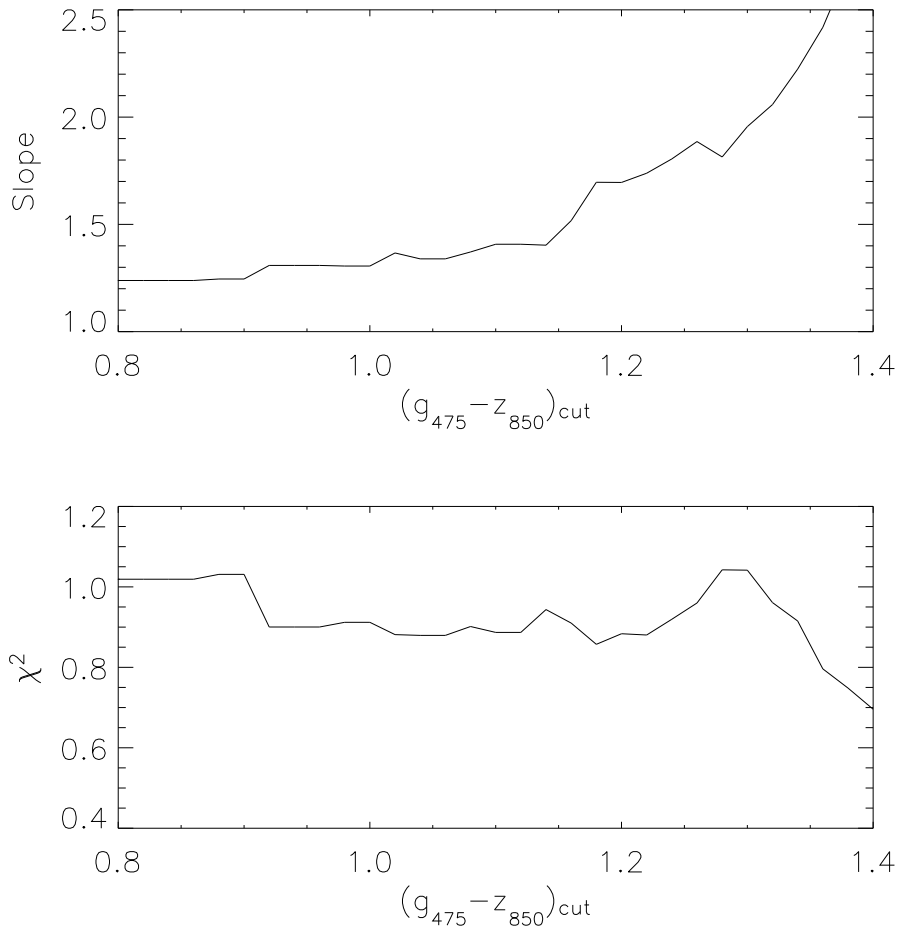}}
\caption {Plot of our calibration sample in different color
ranges $(g_{475}-z_{850})_{cut}~<(g_{475}-z_{850})_0~\le 1.6$. In the top panel, 
the slope obtained by the fit in these different ranges is shown 
as a function of
 $(g_{475}-z_{850})_{cut}$. In the bottom panel, we show the corresponding 
reduced $\chi^2$ for the fits. 
Red galaxies tend to have steeper slopes ($\approx 2$) than the 
slope fitted to the overall sample ($1.3 \pm 0.1$).
 {\label{col_cut}}}
\end{figure}

Calibrations of bluer samples of dwarf early-type galaxies have been
proposed by Jerjen et al.\ (1998, 2000, 2003, 2004) and 
Mieske et al.\ (2003) using ground--based observations of 
$\approx 10$ dwarfs coupled with the predicted color-dependence from
stellar population models.  However, until now, there has not been
a sufficient sample of high-quality SBF measurements in dwarf ellipticals
to permit a dependable {\it empirical} calibration.

Our ACS VCS sample is the most extensive and homogeneous early--type galaxy 
sample available for SBF measurements.  The galaxies span a range in
color of $0.8 \lesssim (g_{475}-z_{850})_0 \lesssim 1.6$~mag, corresponding
to a range in \vi$_0$ between 0.94 and 1.3~mag 
(based on the color transformations from Sirianni et al.\ 2005).
We adopt here the same technique that Tonry et al. (1997) have used for
galaxy groups, fitting the relation:

\begin{equation}\label{eq:cal}
\overline m_{850} =  \langle \overline m^0_{F850LP}\rangle   + \beta [(g_{475}-z_{850})_0 - 1.3]
\end{equation}
for the Virgo Cluster.

Among the 100 galaxies in our full sample, 14 have been omitted from 
our calibration sample because they are edge--on disk or barred galaxies
which are especially challenging for SBF measurements, or because of difficult
sky subtraction.  These galaxies include: VCC~571, VCC~654, VCC~685,  
VCC~1025, VCC~1125, VCC~1192, VCC~1199, VCC~1327,
VCC~1535, VCC~1627, VCC~1720, VCC~1857,  VCC~2048, 
and VCC~2095.
 One galaxy 
in the sample, VCC~1499, has $(g_{475}{-}z_{850})_0 < 0.8$, and is too blue
to be reliably included in a calibration.  There are also several outliers
that do not appear to lie on the average SBF magnitude versus
color relation for the cluster.
Fig.~\ref{allsample} shows the SBF and color measurements for the
remaining 85 galaxies.

To calibrate our SBF measurements in the F850LP filter as a function of the
average galaxy $(g_{475}-z_{850})_0$ color, we calculate average SBF measurements
for each galaxy and fit the relation in Eq.~\ref{eq:cal} to average galaxy
colors, as Tonry et al. (1997) have done for each group in their sample.  All
galaxies included in the fit have an uncertainty on $\overline m_{850}$ less than
0.15~mag (not considering the uncertainty due to the `cosmic' scatter on the SBF
calibration; see above). Errors on average galaxy colors are typically
$\sim\,$0.01~mag. Both uncertainties on magnitude and color are
included in the $\chi^2$ calculations below.

To exclude the outliers from the final calibration, we first make a linear fit
 to Eq.~\ref{eq:cal} to the overall sample
with a robust least absolute deviation technique (Press et al. 1992).
We obtain $\langle\overline m^0_{F850LP}\rangle=29.09 \pm 0.02$~mag and $\beta = 1.25 \pm 0.10$.
We identify four outlier galaxies from iterative 3-$\sigma$ clipping.
The four 3--$\sigma$ outliers are: VCC~21, VCC~538,  VCC~575, and VCC~731. 
Of these, VCC~21 contains numerous bright young star clusters, 
indicative of recent star formation, while VCC~731 (NGC~4365) is known 
from the ground-based SBF survey to be a Virgo member which lies on the far
side of the cluster; we suspect
that the other two may be similar cases.
After excluding these four outliers and the very blue VCC~1499, 
our final sample consists of 82 galaxies.

\begin{figure}
\epsscale{.80}
\centerline{\plotone{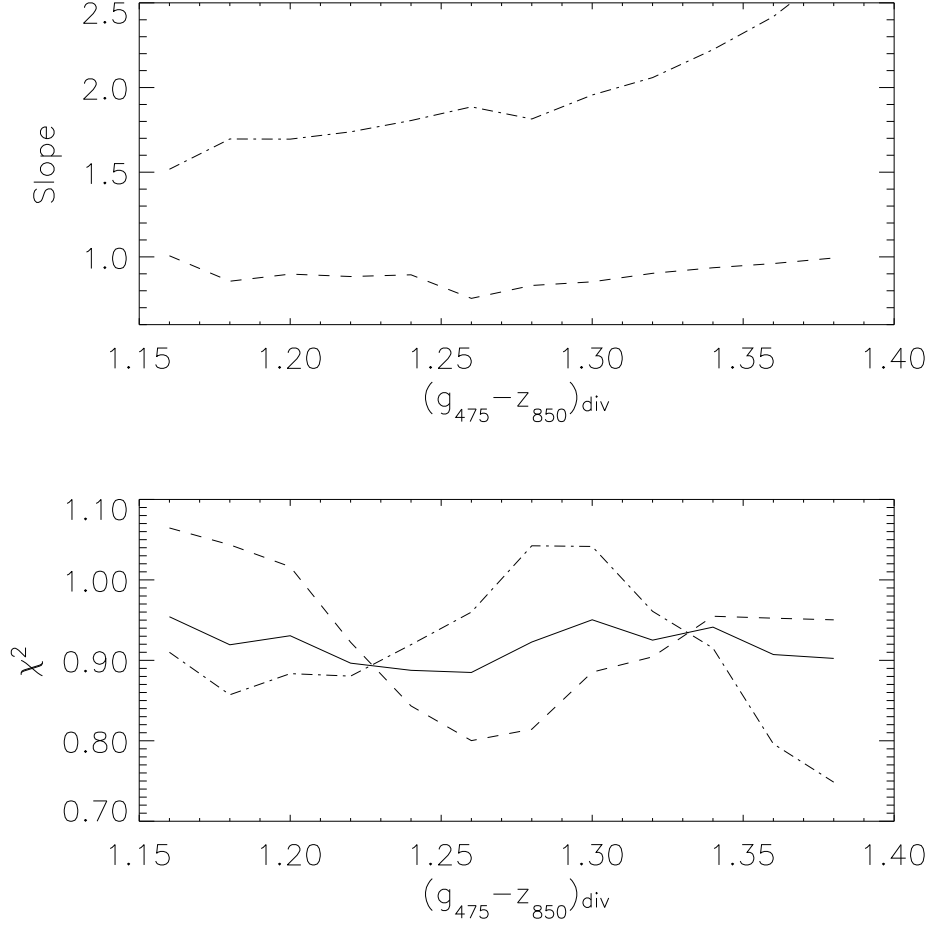}}
\caption {
Best-fit slopes (top panel) and reduced $\chi^2$ values (lower panel)
in two different color ranges 
[blue: $0.9 <(g_{475}-z_{850})_0 \leq (g_{475}-z_{850})_{div}$ and
 red: $(g_{475}-z_{850})_{div} <(g_{475}-z_{850})_0 \le 1.6$] 
are shown as a function of $(g_{475}-z_{850})_{div}$. The dot-dashed 
and dashed lines correspond to the red and blue sample, respectively.
The solid line in the lower panel shows the overall reduced $\chi^2$ for
the full sample.  Red galaxies
(mostly giants) prefer a slope $\approx 2$, while the blue galaxies (mostly
dwarfs) prefer a slope $\approx 1$.
 {\label{col_div}}}
\end{figure}

In Fig.~\ref{slope} our final galaxy sample and an overall fit (the continuous
straight line) are shown. Both uncertainties on magnitude and color have been
included in the fit $\chi^2$ estimation.
We obtain:
\begin{equation}\label{cal}
\overline m_{850} =  (29.09 \pm 0.02) + (1.3 \pm 0.1) \times [(g_{475}-z_{850})_0-1.3]
\nonumber
\end{equation}
in the overall color range  $0.9 \le (g_{475}-z_{850})_0 \le 1.6$.
This color range includes both bright and faint galaxies.
Various lines of evidence have suggested that the slope of the $I$-band
SBF--color relation is steeper for redder galaxies: for instance, the
steep empirical slope found by Ajhar et al. (1997) for
the red centers of bright ellipticals, which was also reproduced by
the reddest composite stellar population models of Blakeslee \etal\
(2001).  Conversely, the few dwarf galaxies in the Tonry \etal\
(1997) sample indicated a significantly shallower slope at the 
blue end, and the SBF measurements for Galactic globular clusters
(Ajhar \& Tonry 1994) showed no dependence on color.  Most recent models
also show a flattening of the relation for bluer populations.

Our sample, more clearly than any previous study, suggests that changes in color range affect the fitted slope, so we have studied the stability of the above relation as a function of the lower color limit, $(g_{475}-z_{850})_{cut}$. When we repeat the fit in different ranges  $(g_{475}-z_{850})_{cut} <(g_{475}-z_{850})_0 \le 1.6$, we obtain the results shown in Fig~\ref{col_cut}. On the top of the figure the fitted slope $\beta$ is shown as a function of  $(g_{475}-z_{850})_{cut}$. On the bottom is shown the corresponding value of $\chi^2$. Red galaxies tend to have slopes that are steeper 
($\approx$ 2) than those obtained by fitting to the overall sample. This
suggests that our sample would be better represented by two distinct samples
of red and blue galaxies, each one having its own calibration slope. 

We fit Eq.~\ref{eq:cal} in two complementary color ranges :  $0.9 <(g_{475}-z_{850})_0  \leq (g_{475}-z_{850})_{div}$ (blue) and  $(g_{475}-z_{850})_{div}  <(g_{475}-z_{850})_0 \le 1.6$ (red), with $1.15 \le (g_{475}-z_{850})_{div} \le 1.35$ (as suggested from  Fig.~\ref{col_cut}).
In Fig.~\ref{col_div} the fitted slopes (top) and the respective $\chi^2$ (bottom) are shown as a function of $(g_{475}-z_{850})_{div}$. Dot-dashed lines correspond to the fit in the color range  $(g_{475}-z_{850})_{div} <(g_{475}-z_{850})_0 \le 1.6$ and dashed lines to the fit in the color range  $0.9 <(g_{475}-z_{850})_0  \leq (g_{475}-z_{850})_{div}$. The continuous line is the total $\chi^2$ from both fits.
The minimum total $\chi^2$ is obtained at $(g_{475}-z_{850})_{div}=1.26$, where the red slope is $1.95 \pm 0.24$ and the blue slope is $0.90 \pm 0.23$.
The two slopes cross at $(g_{475}-z_{850})_{div} \approx 1.3$, which we take to be the color at which the two regimes separate. The $\chi^2$ for the total fit is minimized for both slope and zero-point, 
and the errors in both magnitude and color have been taken into account.

As in Tonry et al. (1997), we find the SBF in S0 and ellipticals to
follow the same trend as a function of $(g_{475}-z_{850})_0$.
Interestingly, as Figs.~2 and 5 show, the dividing color of $(g_{475}-z_{850})_0=1.3$ is
close to that which separates the two most basic galaxy types in
our survey: giant and dwarf early-type galaxies. In fact, in our final sample, 
of 37 galaxies with color $(g_{475}-z_{850})_0 \ge 1.3$, only one is classified
as dwarf. For  $1.2 \le (g_{475}-z_{850})_0 \le 1.3$, 10 galaxies are dwarfs and
8 are giants. For $(g_{475}-z_{850})_0 \le 1.2$, of 26 galaxies, 20 are dwarfs.
Physically then, the two different calibration slopes correspond to
different galaxy types (giants corresponding to the red end, and dwarfs to the
blue end). 

Jerjen et al (2000, 2004) adopted a theoretical calibration to dwarf
elliptical galaxies, based on a two branch prediction in the R--band
SBF versus the $(B-R)_0$ galaxy color from Worthey et al. (1994)
theoretical stellar population models using the Padova isochrones.
While the scatter becomes larger for the bluest dwarfs, we see no
evidence for distinct branches in our large data set; nor do the
Bruzual \& Charlot (2003) models predict such branches in $\overline
z_{850}$ vs $(g_{475}-z_{850})_0$.

\begin{figure}
\epsscale{.80}
\centerline{\plotone{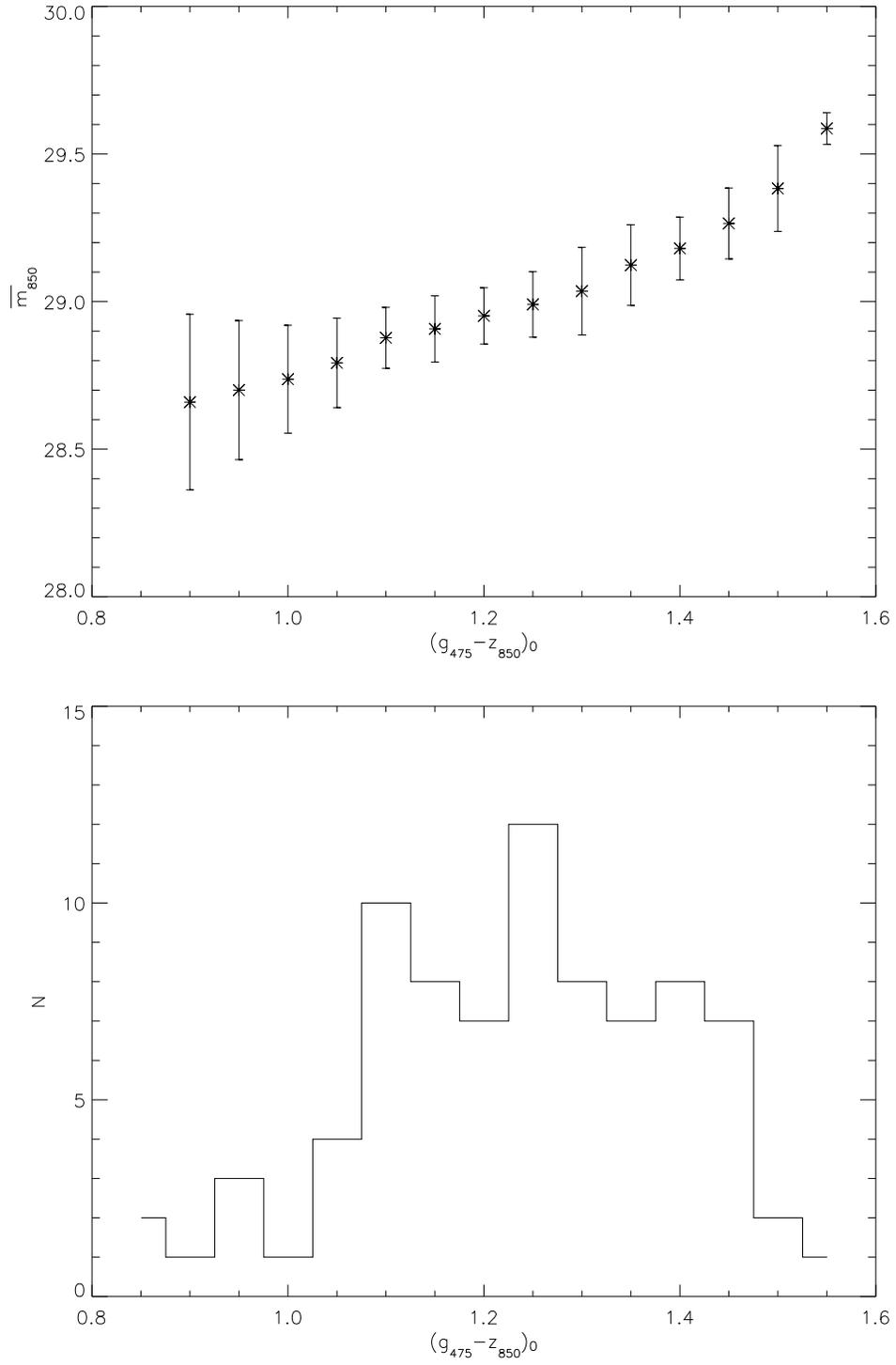}}
\caption {Average and standard deviation of  $\overline m_{850}$ in 0.05~mag color bins (top) and the color histogram (bottom) of our  final calibration sample 
(81 galaxies, plus VCC~21; see text) are shown. {\label{scatter}}}
\end{figure}

Our sample is sufficiently complete at the blue end that we are able
to provide, for the first time in SBF studies, an empirical calibration of SBF
in early--type dwarf galaxies.


Splitting the galaxy sample at $\gz_0 = 1.3$,
we derive the following calibration for ${\overline m}_{850}$: 
\begin{equation}\label{eq:zero}
\overline m_{850} =  29.03 \pm 0.03 + (2.0 \pm 0.2) \times [(g_{475}-z_{850})_0 - 1.3]
\nonumber
\end{equation}
in the range $1.3 <(g_{475}-z_{850})_0 < 1.6$, and
\begin{equation}
\overline m_{850} =   29.03 \pm 0.03  +  (0.9 \pm 0.2) \times [(g_{475}-z_{850})_0 - 1.3]
\nonumber
\end{equation}
in the range $0.9 <(g_{475}-z_{850})_0 \leq 1.3$.
The reduced $\chi^2$ values for the red and blue
subsamples are 0.99 and 0.8, respectively.
The value of the total reduced $\chi^2$ is
calculated as the appropriately weighted sum of 
the $\chi^2$ values from the two independent fits. When
normalized by the total number of objects used in the fits, its value is 0.9.
The reduced $\chi^2$ for the blue subsample is 
lower than that for the red sample. 
This might be due to the difficulty in accounting for the intrinsic 
distance dispersion within Virgo, or to overly conservative error estimates
for the dwarfs. 
One of the outliers rejected above, VCC~21, 
is no longer an outlier with respect to the final, shallower, 
blue-end calibration, and we have added it to the final calibration above, by iteration.

We show in Fig.~\ref{scatter} the average and standard deviation of  $\overline m_{850}$ in $(g_{475}-z_{850})_0$ color bins (top) and the color histogram (bottom) of our  final calibration sample (82 galaxies, plus VCC~21). From the scatter in  $\overline m_{850}$, we are aware that our calibration has a large scatter for colors $(g_{475}-z_{850})_0 < 1$. We will limit our absolute magnitude calibration to the color range  $1.0 < (g_{475}-z_{850})_0 < 1.6$.

We observe a similar behavior of $m_{850}$ as a function of the $(g_{475}-z_{850})_0$
color for the multiple annuli within individual galaxies. However, the precise
value of the slope is less robust and more sensitive to errors in the sky.

\subsection{Calibration Zero-Point}
Empirical calibration of the zero-point of the SBF relation
requires computing the SBF 
absolute magnitude $\overline M_{F850LP}$ from independent 
distance measurements. The ideal situation would be to have 
Cepheid distances for multiple galaxies in our sample,
however there are no galaxies in our sample with Cepheid distances.
We thus calibrate our zero-points to the Virgo Cluster distance modulus
found in the $I$-band SBF survey of Tonry et al.\ (2001), but revised
in accordance with the final set of \hst\ Key Project Cepheid distances
from Freedman \etal\ (2001).
Tonry et al. (2001; Table~4) reported $31.15 \pm 0.03$~mag as
the Virgo SBF distance modulus, where the SBF zero-point 
was calibrated from SBF measurements for individual early-type spiral
galaxies with Cepheid distances given by Ferrarese et al.\ (2000).

However, using the final set of Key Project distances
from Freedman et al. (2001), who adopted the
Udalski et al.\ (1995) period-luminosity relation and added
a metallicity correction, the Tonry et al.\ (2001)
distance modulus for Virgo has to be corrected with a shift of 
$-$0.06~mag, and becomes $31.09 \pm 0.03$~mag
(see discussions by Ajhar et al.\ 2001; Blakeslee et al.\ 2002).
Here, the errorbar reflects the internal error on the mean Virgo
SBF distance; the uncertainties in tying the ground-based SBF
distances to the Cepheid distance scale, and in the
zero-point of the Cepheid distance scale itself, contribute
another $\sim\,$0.15 mag systematic error.
If the metallicity corrections to the Cepheid distances tabulated
by Freedman \etal\ are not applied, then the Tonry et al. (2001) 
Virgo distance modulus would shift instead by $-0.16$~mag 
to $30.99 \pm 0.03$~mag (see Jensen et al.\ 2003).
For our SBF calibration, we will adopt $31.09 \pm 0.03$~mag as 
the Virgo distance modulus.

Thus, scaling the zero-points from Eq.\ref{eq:zero}, we obtain

\begin{equation}
\overline M_{850} =  -2.06 \pm 0.04 + (2.0 \pm 0.2) \times [(g_{475}-z_{850})_0 - 1.3]
\nonumber
\end{equation}
in the range $1.3 <(g_{475}-z_{850})_0 \le 1.6$, and
\begin{equation}
\overline M_{850} =   -2.06 \pm 0.04  +  (0.9 \pm 0.2) \times [(g_{475}-z_{850})_0 - 1.3]
\nonumber
\end{equation}
in the range $1.0 \le (g_{475}-z_{850})_0 \leq 1.3$.

\begin{figure}
\epsscale{.80}
\centerline{\plotone{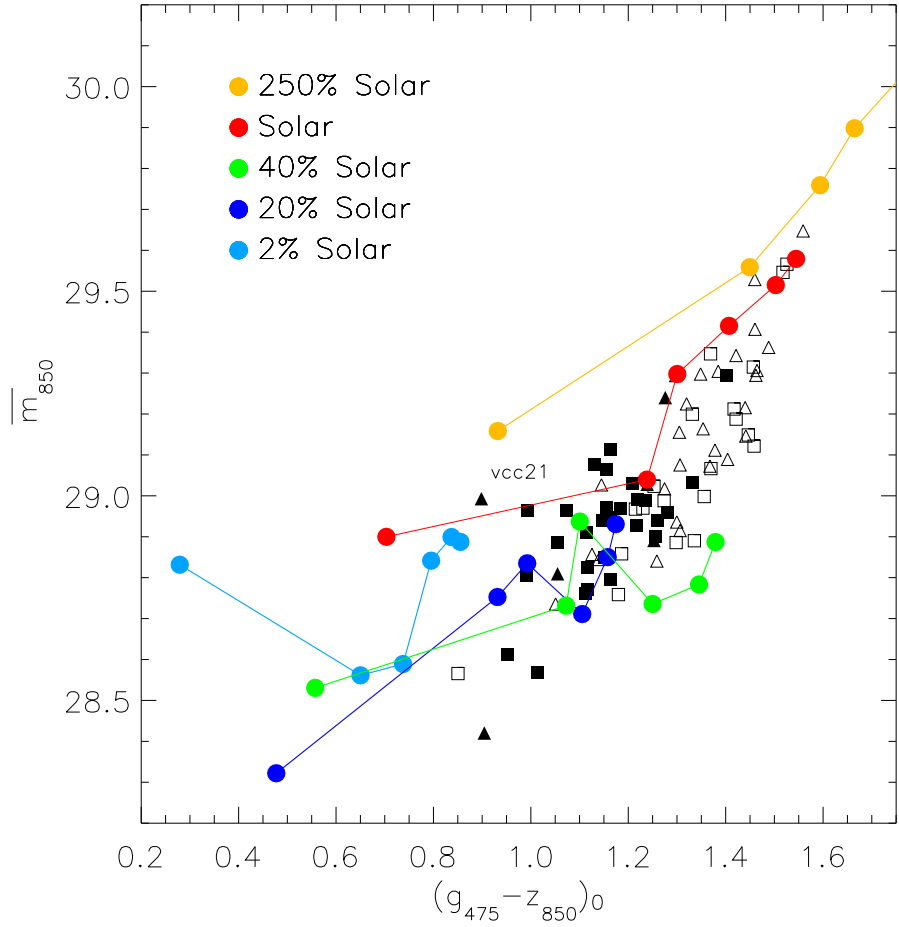}}
\caption {Theoretical predictions from Bruzual \& Charlot (2003)
stellar population models are shown.The red line is for 
solar metallicity, the green for 40\% solar, the blue for 
20\% solar, the light blue for 2\% solar, and the orange for 
metallicity 250\% solar.
The dots on the lines represent, respectively from left to right, ages of 1, 3, 5, 8, 12 and 15~Gyr. {\label{cha}}}
\end{figure}

\section{Predictions from stellar population models}
 
We now compare our results to predictions from the Bruzual \& Charlot (2003) 
stellar population models, which we have calculated from the models and
stellar luminosity functions kindly provided by S.~Charlot.
In Fig.~\ref{cha} we show our data as compared to Bruzual \& Charlot (2003) 
model predictions for different stellar populations.
The lines connect models of the same metallicity: the red line is for 
solar metallicity, the green for 40\% solar, the blue for 
20\% solar, the light blue for 2\% solar, and the orange for 
2.5~times solar metallicity.
The dots on the lines represent, respectively from left to right,
ages of 1, 3, 5, 8, 12 and 15~Gyr.

The Bruzual \& Charlot models predict $\overline M_{850}$. 
To compare these predictions with the data, we add to the
theoretical $\overline M_{850}$
a distance modulus such that the predicted average $\overline m_{850}$ for 
solar and 40\% solar metallicity at $(g_{475}-z_{850})_0=1.2$~mag
is equal to  the average observed $\overline m_{850}$ at 
 the same $(g_{475}-z_{850})_0$ 
for our calibration sample. This procedure yields a theoretically
predicted SBF Virgo Cluster distance modulus of
$\overline M_{850}-\overline m_{850} = 30.94$~mag, which is
entirely independent of the Cepheid distance scale, but
we caution that it depends on the color range we
have chosen for shifting the models into agreement with the data.

This theoretically calibrated SBF Virgo Cluster distance modulus
differs by $\approx$0.15~mag with respect to the revised 
Tonry et al.\ (2001) Virgo distance modulus of 31.09 mag adopted above.
However, the two are consistent within
the uncertainties from the zero-point of the Cepheid distance scale,
the anchoring of the ground-based SBF measurements to the Cepheids, 
and the various uncertainties in the stellar population models
(see Blakeslee et al.\ 2001 for a discussion).  It is also worth noting 
that if we had adopted the Virgo distance modulus of $30.99 \pm 0.03$
obtained by calibrating SBF to the Freedman \etal\ (2001) Cepheid distances
that use the new period--luminosity calibration but omit
the metallicity correction, then the difference
with respect to the model-predicted distance modulus would be
only $0.05$~mag.

Our sample of early-type galaxies spans a wide range in metallicity
and, presumably, age.   Like the observations,
the theoretical predictions also show a steeper slope
at the red end, and a shallower at the blue end.
The slopes predicted from linear fits to the overall model predictions
 are $0.0 \pm 0.5$ for  $0.9 <(g_{475}-z_{850})_0 \leq 1.25$~mag,
$2.6\pm 0.9$ for  $1.25 <(g_{475}-z_{850})_0 \leq 1.6$~mag, and $1.26 \pm 0.3$
for $0.9 <(g_{475}-z_{850})_0 \leq 1.6$~mag
(in very good agreement with our empirical
slope over the full range). 
If we fit only metallicities less than or equal to solar,
at the red end the slope will be $2\pm 1$, compared to $0.4 \pm 0.4$
in the blue.
The agreement between theoretical predictions
and our empirical calibration is somewhat better at the red end. 
This is likely due to better accuracy in the modeling of
highly evolved populations.  The SBF predictions for young to 
intermediate-age populations is less certain due mainly to the uncertainties
in the modeling of the asymptotic giant branch contributions.
Furthermore, models of old stellar populations are better
constrained by observations of Galactic globular clusters.
For these reasons, we prefer our empirical blue-end slope,
although additional SBF measurements of large homogeneous samples of
dwarf galaxies (e.g., in the Fornax cluster) would be helpful
in refining the SBF calibration for blue early-type galaxies.

Considering the uncertainty in the Virgo distance modulus, we
conclude that our $z_{850}$ SBF data for the giant early-type 
galaxies are well reproduced by the models for stellar populations
ages greater than 3~Gyr and metallicities in the range between
40\% and 250\% solar.  The data for the dwarf galaxies are
consistent with either much younger (less than 3~Gyr) solar
metallicity populations or intermediate-to-old stellar
populations with metallicities in the range 20\%--40\% solar.

\section{Conclusions}

As part of the ACS Virgo Cluster Survey, we have observed 100 early--type
galaxies in the Virgo Cluster.  We have measured SBF magnitudes in
these galaxies with the aim of determining their distances and
the three-dimensional structure of Virgo.
We have presented the first calibration of the SBF method using
the ACS instrument, giving the F850W SBF magnitude 
as a function of galaxy $(g_{475}-z_{850})_0$ color.
Obvious outliers, most likely due to a real dispersion in
distance for galaxies in the Virgo Cluster region,
have been omitted from this calibration,
leaving 82 galaxies with  $(g_{475}-z_{850})_0$
colors ranging between 0.9 and 1.6~mag.
Over this color range, the blue (mainly dwarf) and red (mainly giant) galaxies 
follow different linear relations, with the slope being shallower
at the blue end. For the first time in SBF studies, our sample allows us
to empirically calibrate SBF for dwarf early--type galaxies. 

By comparing with stellar population models from Bruzual \& Charlot (2003),
we find that the 
slope of the model prediction for $\overline m_{850}$ with \gz$_0$ color is
highly consistent with the observed relation for the red galaxy subsample.
There is also general consistency at the blue end, though
the stellar population models with young ages and 
low-metallicities predict a somewhat shallower slope than observed.
This underscores the importance of empirical calibrations,
as well as more realistic composite 
populations in any effort to model the behavior of the SBF method
theoretically, especially for dwarf galaxies, for which the scatter in
stellar population properties is larger. 
%
A forthcoming paper will present the distances to the individual 
galaxies derived with this calibration, and examine the
issue of Virgo's three dimensional structure and its implications for
galaxy and globular cluster properties.


\acknowledgments
We thank St\'ephane Charlot for providing theoretical stellar population models in the ACS filters 
and Gerhard Meurer for useful discussions.
Support for program GO-9401 was provided through a grant from the Space
Telescope Science Institute, which is operated by the Association of 
Universities for Research in Astronomy, Inc., under NASA contract NAS5-26555. 
ACS was developed under NASA contract NAS 5-32865.
S.M. and J.P.B. acknowledge additional support from NASA grant 
NAG5-7697 to the ACS Team.
P.C. acknowledges support provided by NASA LTSA grant NAG5-11714.
M.J.W. acknowledges support through NSF grant AST-0205960.
D.M. acknowledges support provided by NSF grants AST-0071099, AST-0206031, 
AST-0420920 and AST-0437519, by NASA grant NNG04GJ48G, and by grant 
HST-AR-09519.01-A from STScI. 
M.M. acknowledges support from the Sherman M. Fairchild foundation. 
This research has made use of the NASA/IPAC Extragalactic Database (NED)
which is operated by the Jet Propulsion Laboratory, California Institute
of Technology, under contract with the National Aeronautics and Space Administration. 


\newpage


\clearpage


\begin{thebibliography}{}

\bibitem[Ajhar et al. 1997]{aj97}
Ajhar E.A., Lauer, T.R., Tonry, J.L. et al., 1997, AJ, 114, 626

\bibitem[Ajhar and Tonry 1994]{aj97}
Ajhar E.A. \& Tonry, J.L. 1994, ApJ, 429, 557

\bibitem[Ajhar et al. 2001]{ajh01}
 Ajhar E.A., Tonry, J.L.,  Blakeslee, J.P. et al. 2001, ApJ, 559, 584
 

\bibitem[Benitez et al. 2004]{ben04}
Benitez, N., Ford, H. Bowens, R. et al. 2004, ApJS, 150, 1

\bibitem[Blakeslee  \& Tonry 1995]{bla95}
Blakeslee J.P. \& Tonry, J.L., 1995, ApJ, 442, 579;

\bibitem [Blakeslee et al. 1999]{bla99}
Blakeslee, J. P., Ajhar, E. A., Tonry, J. L., 1999, in Post-Hipparcos
Cosmic Candles, eds. A. Heck \& F. Caputo (Boston: Kluwer), 181



\bibitem[Blakeslee et al. 2001]{bla01}
Blakeslee, J.P., Vazdekis, A., \& Ajhar, E.A.\ 2001, \mnras, 320, 193 

\bibitem [Blakeslee et al. 2002]{bla02}
Blakeslee, J.P., Lucey, J.R., Tonry, J.L. et al. 2002, MNRAS, 330, 443


\bibitem[Bruzual \& Charlot(2003)]{bru03} Bruzual, G.~\& 
Charlot, S.\ 2003, \mnras, 344, 1000 



\bibitem[C\^ot\'e et al. 2004]{cot04}
C\^ot\'e, P., Blakeslee, J.P., Ferrarese, L., Jord\'an, A., Mei, S., Merritt, D., Milosavljevi\'c, M.,
Peng, E.W., \& West, M.J. 2004, \apjs, 153, 223 (Paper~I)

\bibitem[Ferrarese et al. 2000]{fer00a}
Ferrarese L., Mould, J.R., Kennicutt R.C. 2000, ApJ, 529, 745




\bibitem[Ferrarese et al. 2003]{fer03}
Ferrarese, L., C\^ot\'e, P. \& Jord\'an, A.  2003, ApJ, 599, 1302

\bibitem[Ford et al. 1998]{for98}
Ford, H.C. et al. 1998, Proc. SPIE, 3356, 234

\bibitem[Freedman et al. 2001]{fre01}
Freedman et al. 2001,\apj, 553, 47

\bibitem[Fruchter \& Hook 2002]{fru02} 
Fruchter, A.~S.~\& Hook, R.~N.\ 2002, \pasp, 114, 144 


\bibitem[Harris 1991]{har91}
Harris, W.E. 1991, ARA\&A, 29, 543;
 

\bibitem[Jacoby et al. 1992]{jac92}
Jacoby, G.H., Branch, D., Ciardullo, R., Davies, R.L., Harris, W.E., Pierce, M.J., Pritchet, C.J., Tonry, J.L., Welch, D.L. 1992, PASP, 104, 599



\bibitem[Jensen et al. 1999]{jen99}
Jensen, J.B., Tonry, J.L., Luppino, G.A., 1999, ApJ, 510, 71



\bibitem[Jensen et al. 2003]{jen03} Jensen, J.~B., Tonry, 
J.~L., Barris, B.~J., Thompson, R.~I., Liu, M.~C., Rieke, M.~J., Ajhar, 
E.~A., \& Blakeslee, J.~P.\ 2003, \apj, 583, 712 



\bibitem[Jerjen  2003]{jer03}
Jerjen, H. 2003, A\&A, 398, 63


\bibitem[Jerjen et al.  1998]{jer98} 
Jerjen, H., Freeman, K.C., \& Binggeli, B. 1998, \aj, 116, 2873 

\bibitem[Jerjen et al.  2000]{jer00} 
Jerjen, H., Binggeli, B., \& Freeman, K.C. 2000, \aj, 119, 166 


\bibitem[Jerjen et al.  2004]{jer04} 
Jerjen, H., Binggeli, B., \& Barazza, F.~D. 2004, \aj, 127, 771 

\bibitem[Jord\'an et al. 2004]{jor04}
Jord\'an, A., Blakeslee, J.P., Peng, E.W., Mei, S., C\^ot\'e, P., Ferrarese, L., Tonry, J.L., Merritt, D., Milosavljevi\'c, M., \& West, M.J. 2004, ApJS, 154, 509 (Paper~II).

\bibitem[Liu \& Graham 2001]{liu01}
Liu, M.C. \& Graham J.R. 2001, ApJL, 557, 31

\bibitem[Liu et al. 2002]{liu02} 
Liu, M.~C., Graham, J.~R., \& Charlot, S. 2002, \apj, 564, 216 



\bibitem[Mei et al. 2001]{mei01} 
Mei, S., Silva, D.~R., \& Quinn, P.~J. 2001, \aap, 366, 54 



\bibitem[Mei et al.  2003]{mei03}
Mei, S.,  Scodeggio, M.,  Silva, D.R, Quinn, P.J. 2003, A\&A,  399, 441

\bibitem[Mei et al. 2005]{mei05}
Mei, S., Blakeslee, J.P., Jord\'an, A., Peng, E.W., C\^ot\'e, P., Ferrarese, L., Tonry, J.L., Merritt, D., Milosavljevi\'c, M., \& West, M.J. 2005, ApJS, in press (Paper~IV).



\bibitem[Mieske et al. 2003]{mie03b}
 Mieske, S., Hilker, M., \& Infante, L.\ 2003, \aap, 403, 43 


\bibitem[Neilsen \& Tsvetanov 2000]{nei00b}
Neilsen, E.H. \& Tsvetanov, Z.I. 2000, \apj, 536, 255


\bibitem[Pahre \& Mould 1994]{pah94}
Pahre, M.A. \& Mould J.R. 1994, ApJ, 433, 567





\bibitem[Press et al. 1992]{press2}
 Press, W.H. et al. 1992,
{\it Numerical Recipes},  Cambridge University Press, New York

\bibitem[Schlegel et al. 1998]{sch98}
Schlegel D.J., Finkbeiner D.,P., Davis M. 1998, ApJ,500,525
 
\bibitem[Sirianni et al. 2005]{si05}
Sirianni, M. {\it et al.} 2005, PASP, submitted


\bibitem[Tonry \& Schneider 1988]{ts88}
Tonry, J.L. \& Schneider, D.P., 1988, AJ, 96, 807


\bibitem[Tonry et al. 1997]{ton97}
Tonry, J.L., Blakeslee, J.P., Ajhar, E.A. {\it et al.} 1997, ApJ, 475, 399

\bibitem[Tonry et al. 2000]{ton00}
Tonry, John L., Blakeslee, J.P., Ajhar, Edward A. {\it et al.} 2000, ApJ, 530, 625

\bibitem[Tonry et al. 2001]{ton01}
Tonry, J.L., Dressler,A.,   Blakeslee, J.P.  {\it et al.} 2001, ApJ, 546, 681 

\bibitem[Udalski et al. 1999]{uda99}
Udalski, A., Soszynski, I., Szymanski, M.  {\it et al.} 1999, Acta Astron., 49, 223

\bibitem[West \& Blakeslee 2000]{we00}
West, M.J. \& Blakeslee, J.P. 2000, ApJL, 543, 27


\bibitem[Whitmore et al. 1995]{whi95}
Whitmore, B.~C., Sparks, W.~B., Lucas, R.~A., Macchetto, F.~D., \& Biretta, J.~A.\ 1995, 
\apjl, 454, L73 

\bibitem[]{wo94}
Worthey, G., 1994, ApJS, 95, 107


\end{thebibliography}
\end{document}